\documentclass[lettersize,journal]{IEEEtran}
\usepackage{amsmath,amsfonts}
\usepackage{algorithmic}
\usepackage{array}
\usepackage[caption=false,font=normalsize,labelfont=sf,textfont=sf]{subfig}
\usepackage{textcomp}
\usepackage{stfloats}
\usepackage{url}
\usepackage{verbatim}
\usepackage{graphicx}
\def\BibTeX{{\rm B\kern-.05em{\sc i\kern-.025em b}\kern-.08em
    T\kern-.1667em\lower.7ex\hbox{E}\kern-.125emX}}
\usepackage{balance}

\usepackage{cite}
\usepackage{tikz}
\usepackage{pgfplots}
\usepackage[nolist]{acronym}
\usepackage{amssymb}
\usetikzlibrary{calc}
\usepackage{color} 
\usepackage{units}

\usepackage{fancyhdr}

\usepackage{graphicx}

\usepackage{textcomp}

\begin{acronym}
	\acro{AWGN}{Additive White Gaussian Noise}
	\acro{CDF}{Cumulative Distribution Function}	
	\acro{CP}{Cyclic Prefix}	
	\acro{OFDM}{Orthogonal Frequency Division Multiplexing}
	\acro{DVB}{Digital Video Broadcasting}
	\acro{WLAN}{Wireless Local Area Network}
	\acro{LTE}{Long Term Evolution}
	\acro{ISI}{Inter Symbol Interference}
	\acro{IFFT}{Inverse Fast Fourier Transform}
	\acro{FFT}{Fast Fourier Transform}
	\acro{ICI}{Inter-Carrier Interference}
	\acro{WSSUS}{Wide-Sense Stationary Uncorrelated Scattering}
	\acro{pdf}{probability density function}
	\acro{QAM}{Quadrature Amplitude Modulation}
	\acro{PAM}{Pulse-Amplitude Modulation}
	\acro{PSK}{Phase-Shift Keying}
	\acro{LS}{Least Squares}
	\acro{2D}{two-dimensional}
	\acro{MMSE}{Minimum Mean Squared Error}
	\acro{LMMSE}{Linear Minimum Mean Squared Error}
	\acro{MSE}{Mean Squared Error}
	\acro{BER}{Bit Error Ratio}
	\acro{BEP}{Bit Error Probability}
	\acro{LTV}{Linear Time Variant}
	\acro{WSS}{Wide-Sense Stationary Uncorrelated Scatterers}
	\acro{SIR}{Signal-to-Interference Ratio}
	\acro{SINR}{Signal-to-Interference plus Noise Ratio}
	\acro{SNR}{Signal-to-Noise Ratio}
	\acro{PACE}{Pilot-symbol-Aided Channel Estimation}
	\acro{RX}{receive}
	\acro{ML}{Maximum Likelihood}
	\acro{GPS}{Global Positioning System}
	\acro{FBMC}{Filter Bank Multi-Carrier}
	\acro{MIMO}{Multiple-Input and Multiple-Output}
	\acro{SISO}{Single-Input and Single-Output}	
	\acro{UFMC}{Universal Filtered Multi-Carrier}
	\acro{OQAM}{Offset Quadrature Amplitude Modulation}
	\acro{ZF}{Zero Forcing}
	\acro{RMS}{Root Mean Square}
	\acro{LO}{Local Oscillator}
	\acro{TX}{Transmitter}
	\acro{RX}{Receiver}
	\acro{BS}{Base Station}
	\acro{OTA}{Over The Air}
	\acro{TRP}{Transmission Reception Point}
	\acro{TRX}{Transceiver}
	\acro{RF}{Radio Frequency}
	\acro{HW}{Hardware}
	\acro{DL}{Downlink}
	\acro{UL}{Uplink}
	\acro{TDD}{Time Division Duplex}
	\acro{SRS}{Sounding Reference Signals}
	\acro{DMRS}{Demodulation Reference Signals}
	\acro{UE}{User Equipment}
	\acro{PA}{Power Amplifier}
	\acro{LF}{Low Frequency}
\end{acronym}

\definecolor{mycolor}{rgb}{0, 0, 0.9333}

\makeatletter
\def\ps@IEEEtitlepagestyle{%
	\def\@oddfoot{\mycopyrightnotice}%
	\def\@oddhead{\mycopyrightnoticeA}
	\def\@evenhead{\mycopyrightnotice}
	\def\@evenfoot{\mycopyrightnoticeA}%
}
\def\mycopyrightnotice{%
	\begin{minipage}{\textwidth}
		\centering \scriptsize 
		\copyright 2022 IEEE. Personal use is permitted, but republication/redistribution requires IEEE permission.
		See https://www.ieee.org/publications/rights/index.html for more information.
	\end{minipage}
}
\def\mycopyrightnoticeA{%
	\begin{minipage}{\textwidth}
		\centering \scriptsize 
	This article has been accepted for publication in IEEE Communications Letters. This is the author's version which has not been fully edited and
	content may change prior to final publication. The final version of record is available at:	\color{mycolor}{https://doi.org/10.1109/LCOMM.2022.3189968}
	\end{minipage}
}
\makeatother

\begin{document}

\title{Correctly Modeling TX and RX Chain in (Distributed) Massive MIMO - New Fundamental Insights on Coherency}

\author{Ronald Nissel 
\thanks{Manuscript received April 8, 2022; revised May 24, 2022; accepted July 7, 2022. The associate editor coordinating the review of this letter and approving it for publication was P. Nguyen.}
\thanks{	
	Ronad Nissel is with Huawei Technologies Sweden AB, Gothenburg, Sweden
	(e-mail: ronald.nissel@huawei.com).}
}

\markboth{IEEE COMMUNICATIONS LETTERS}{asdf}

\maketitle

\begin{abstract}
This letter shows that the TX and RX models commonly used in literature for downlink (distributed) massive MIMO are inaccurate, leading also to inaccurate conclusions. In particular, the \ac{LO} effect should be modeled as $+\varphi$ in the transmitter chain and $-\varphi$ in the receiver chain, i.e., different signs. A common misconception in literature is to use the same sign for both chains. 
By correctly modeling TX and RX chain, one realizes that the \ac{LO} phases are included in the reciprocity calibration and whenever the \ac{LO} phases drift apart, a new reciprocity calibration becomes necessary (the same applies to time drifts). 
Thus, free-running \ac{LO}s and the commonly made assumption of perfect reciprocity calibration (to enable blind DL channel estimation) are both not that useful, as they would require too much calibration overhead.
Instead, the \ac{LO}s at the base stations should be locked and relative reciprocity calibration in combination with downlink demodulation reference symbols should be employed.

\end{abstract}

\begin{IEEEkeywords}
Massive MIMO, Phase Noise, Downlink, Time-Division Duplex (TDD), Hardware, Distributed MIMO
\end{IEEEkeywords}

\section{Introduction}
\acresetall 
\IEEEPARstart{M}{assive} \ac{MIMO} is one of the most important key technologies in 5G~\cite{5G} and expected to stay important in all future mobile systems. Moreover, distributed massive \ac{MIMO} might become more and more important, i.e., coherent joint transmission/reception not only within the antenna array of one \ac{TRP}, but also between \ac{TRP}s.
To keep the pilot overhead relatively low, Massive \ac{MIMO} employs \ac{TDD}, allowing to estimate the \ac{DL} channel based on \ac{UL} \ac{SRS}. However, for this to work, reciprocity calibration becomes necessary, i.e., differences in \ac{TX} and \ac{RX} chain need to be calibrated out. 
A common misconception in literature seems to be that reciprocity calibration only calibrates out the ``obvious'' \ac{HW} elements, i.e., S21 of filters, power amplifiers, etc. and thus is only needed once every few hours \cite{bjornson2017massive}.
However, this is not accurate, reciprocity calibration also includes the \ac{LO} phases and any timing offsets. Whenever \ac{LO} phases drift apart, a new reciprocity calibration becomes necessary. The same applies to drifts in timing offsets. Thus, the recommendation frequently mentioned in recent papers to use free-running \ac{LO}s is not very practical, as it would cause too much calibration overhead (among other things).

The influence of \ac{LO} phase drifts in (distributed) massive \ac{MIMO} was investigated in \cite{khanzadi2015capacity,krishnan2015linear, jiang2021impact,bjornson2015distributed}, under the inaccurate assumption that the \ac{LO} effect can be modeled as  $+\varphi$ for both, the \ac{TX} and the \ac{RX} chain, leading to the inaccurate conclusion that any \ac{LO} phase drift will implicitly be estimated by \ac{SRS} and thus free-running \ac{LO}s can be useful.
By employing a correct model, i.e., $+\varphi$ for the \ac{TX} and $-\varphi$ for the \ac{RX} chain, on the other hand, one realizes that coherent joint transmission requires locked \ac{LO}s, either directly by ``cables'' or indirectly by continuously sending reciprocity calibration signals. 
The paper in \cite{bjornson2015distributed} employs a (mostly) correct  model in Section~2, an implicit result of defining channel reciprocity as ``Hermitian'' instead of the more common ``transposed'', but in Section~3 the paper uses the one-sided effective channel and thus an inaccurate model.
The paper in \cite{papazafeiropoulos2016impact} employs a similar (mostly correct) model as \cite{bjornson2015distributed} and identifies some fundamental issues with it, i.e., the rate becomes zero for blind \ac{DL} channel estimation because of the random \ac{LO} phases. The paper solves this challenge by setting the \ac{LO} phases to zero at time $n=0$,
thus implicitly assuming perfect reciprocity calibration at time $n=0$, but without mentioning it or discussing the practical implications.
In general, none of the papers in \cite{khanzadi2015capacity,krishnan2015linear, jiang2021impact,bjornson2015distributed,papazafeiropoulos2016impact} pointed out the relationship between \ac{LO} phase drifts and reciprocity calibration.

Coherent signal detection requires the \ac{UE} to estimate the effective \ac{DL} channel.
Many papers \cite{bjornson2017massive,ngo2017no} assume perfect reciprocity calibration to enable simple blind \ac{DL} channel estimation at the \ac{UE}, i.e., estimation without \ac{DL} pilots.
By correctly modeling  \ac{TX} and \ac{RX} chain, however, one realizes that any \ac{LO} phase drift between \ac{UE} and base station would also require an update of the \ac{UE} reciprocity calibration factor, which in turn causes significant overhead and is one of the reasons why the assumption of perfect reciprocity calibration is not very practical. Instead, a better solution is to use relative reciprocity calibration in combination with \ac{DMRS}.

The findings in this paper will not change the way how real massive \ac{MIMO} networks are built \cite{5G,shepard2012argos}, as they already follow the proposed guidelines (lock \ac{LO}s, send as many \ac{DMRS} as needed), but rather provide theoretical justifications for those well-known guidelines and to clarify some common misunderstandings in literature.

To support reproducibility, the Python code used in this paper can be downloaded at \textbf{\url{https://github.com/rnissel}}.

\section{Correctly Modeling TX and RX Chain}

\begin{figure}[!t]
	\centering
	\includegraphics[width=\columnwidth]{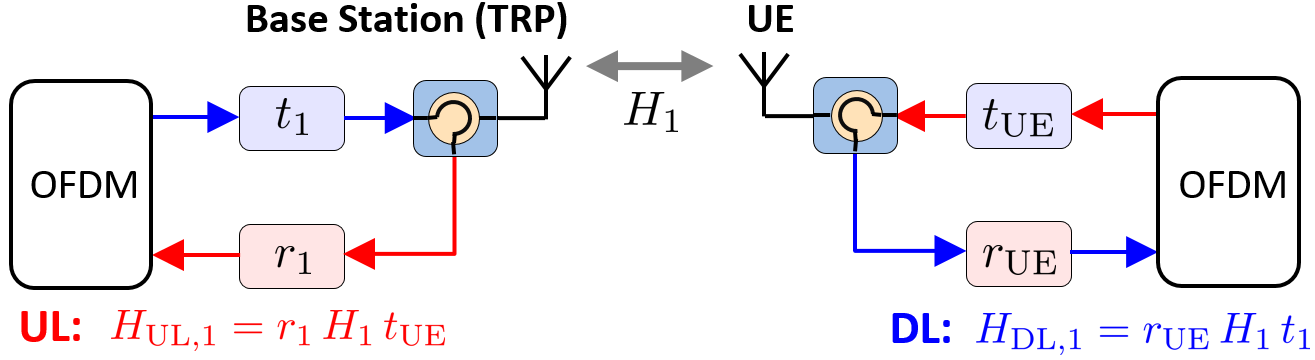}
	\caption{Reciprocity only holds for propagation channel $H_1$, but the UL channel and DL channel are different because of differences in TX and RX.}
	\label{fig:SimpleBlockDiagram}
\end{figure}

\begin{figure}[!t]
	\centering
	\includegraphics[width=7cm]{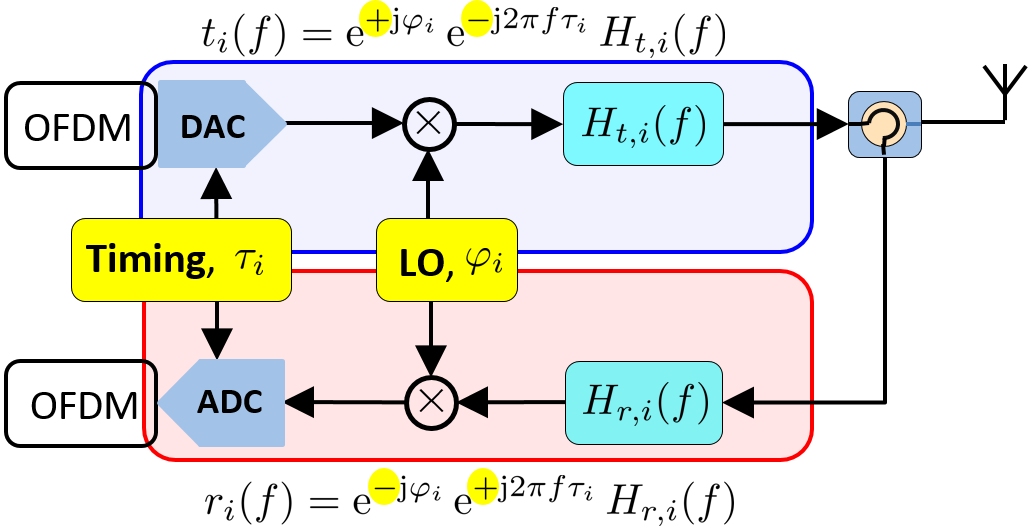}
	\caption{The LO effect on the TX chain is $+\varphi$ while for the RX chain it is $-\varphi$. A similar relationship exists for timing errors.}
	\label{fig:TXRXchain}
\end{figure}

\noindent Figure~\ref{fig:SimpleBlockDiagram} shows a simplified block diagram of a \ac{TDD} transmission system. 
In \ac{UL} direction, the signal goes through \ac{UE} \ac{TX} chain $t_\text{UE}$, reciprocal propagation channel  $H_1$ (which includes everything between the two circulators, i.e., also antenna, feeding network, RF filters etc.) and base station \ac{RX} chain $r_1$. The \ac{DL} can be described in a similar way, so that \ac{UL} channel $H_{\text{UL},1}$ and \ac{DL} channel $H_{\text{DL},1}$ can be written as:
\begin{align}
	\label{eq:H_UL}
	H_{\text{UL},1} &= r_1 \, H_1 \, t_\text{UE}
	\\
	H_{\text{DL},1} & = r_\text{UE} \, H_1 \, t_1
	.
	\label{eq:H_DL}
\end{align}
The \ac{TX} and \ac{RX} chains can be further decomposed into:
\begin{align}
	\label{eq:tx}
	t_i(f) &= \mathrm{e}^{+\mathrm{j} \varphi_{i}} \,\mathrm{e}^{ -\mathrm{j} 2 \pi f \tau_{i}} \, H_{t,i}(f)
	\\
	\label{eq:rx}
	r_i(f) &= \mathrm{e}^{ -\mathrm{j} \varphi_{i}} \,\mathrm{e}^{ +\mathrm{j} 2 \pi f \tau_{i}} \, H_{r,i}(f) 
	,
\end{align}
as also illustrated in Figure~\ref{fig:TXRXchain} and formally proven in the appendix. 
It is important to emphasize here that $t_i$ and $r_i$ are mathematical constructs and only the product $r_j t_i$ has real physical meaning. 
Note that \eqref{eq:tx} and \eqref{eq:rx} provide a general description of the $i$-th \ac{TRX} and can be used to describe the \ac{TRX} at the base station but also at the \ac{UE}, though for clarity the subscript ``UE'' is used for the latter.
Moreover, $t_i(f)$ and $r_i(f)$ now include a frequency dependency $f$ because of \ac{OFDM} \cite{NisselDissertation}, where subcarriers at frequency $f=Fl$ are transmitted in parallel, with $F$ denoting the subcarrier spacing and $l$ the subcarrier index, as explained in more detail in the appendix.
To simplify the explanation, however, the frequency dependency will be ignored from now on, but one must keep in mind that there are many more subcarriers.
The variable $\varphi_{i}$ in \eqref{eq:tx} and \eqref{eq:rx} describes the \ac{LO} phase,  $\tau_{i}$ the timing offset\footnote{
	the +sign in the \ac{RX} chain is caused by delaying the demodulation process itself, not the signal, i.e., $\int_{\tau_r}^{\tau_r + {1}/{F}} 
	\! \! \!
	s(\text{t})
	g^*_l(\text{t} - \tau_r)
	\mathrm{d}\text{t} = 	\int_{0}^{{1}/{F}} 
	s(\text{t} + \tau_r)
	g^*_l(\text{t})
	\mathrm{d}\text{t}$, see also the appendix for details.
} and $H_{t,i}$ includes the ``obvious'' \ac{HW} elements of the transmitter chain, such as power amplifiers and filters, but also the less obvious elements, such as the additional delay from the common \ac{LO} phase $\varphi_{i}$ to the actual \ac{LO} phase used for up-conversion, i.e., $\varphi_{t,i} = \varphi_{i} + \Delta \varphi_{t,i}$. The variable $H_{r,i}$ is similar to $H_{t,i}$, but for the \ac{RX} chain.
Typically $H_{r,i}$ and $H_{t,i}$ will be relatively stable over time (e.g., hours, depending mainly on temperature variations) while $\varphi_{i}$ and $\tau_{i}$ might change very fast, especially for free-running \ac{LO}s (e.g., $100\,\text{µs}$, depending of course on \ac{HW} quality and implementation).

The key difference of \eqref{eq:tx} and \eqref{eq:rx} compared with the \ac{TX} and \ac{RX} model of other papers \cite{khanzadi2015capacity,krishnan2015linear, jiang2021impact,bjornson2015distributed} is the different sign in the \ac{RX} chain (and the inclusion of time offsets):
\begin{align}
	\label{eq:wrongRight}
	\begin{matrix}
		\textbf{Inaccurate} \, \text{[3]-[6]}   & & & & & \textbf{Correct}\\
		t_i \propto \mathrm{e}^{+\mathrm{j} \varphi_{i}}  & & & & &	t_i  \propto \mathrm{e}^{+\mathrm{j} \varphi_{i}} \\
		r_i \propto \mathrm{e}^{+\mathrm{j} \varphi_{i}}  & & & & &	r_i \propto \mathrm{e}^{-\mathrm{j} \varphi_{i}} 
	\end{matrix}
.
\end{align}
The formal proof for \eqref{eq:wrongRight} can be found in the appendix, while in the following let us consider a more intuitive explanation. For the sake of argument, all other components except the \ac{LO} are ignored:
\begin{itemize}
	\item Suppose \ac{TX} and \ac{RX} of the same \ac{TRX} are directly connected together. One would expect that the transfer function becomes ``1''  because up-conversion and down-conversion perfectly cancels each other and there are no other components in this example. The only way how this can happen is by $\mathrm{e}^{-\mathrm{j} \varphi}\mathrm{e}^{+\mathrm{j} \varphi}=1$, i.e., the \ac{LO} influence on \ac{TX} and \ac{RX} chain must have different signs.
	\item An absolute phase does not exist, one can only define relative phases. Thus, the transmission from \ac{TX} to \ac{RX} must be invariant to any common phase transformation of the form $\varphi \rightarrow \varphi' + b$, which is only possible if the \ac{LO} effect on \ac{TX} and \ac{RX} have different signs.
\end{itemize}

\section{Implications of TX and RX Model}

\begin{figure}[!t]
	\centering
	\includegraphics[width=\columnwidth]{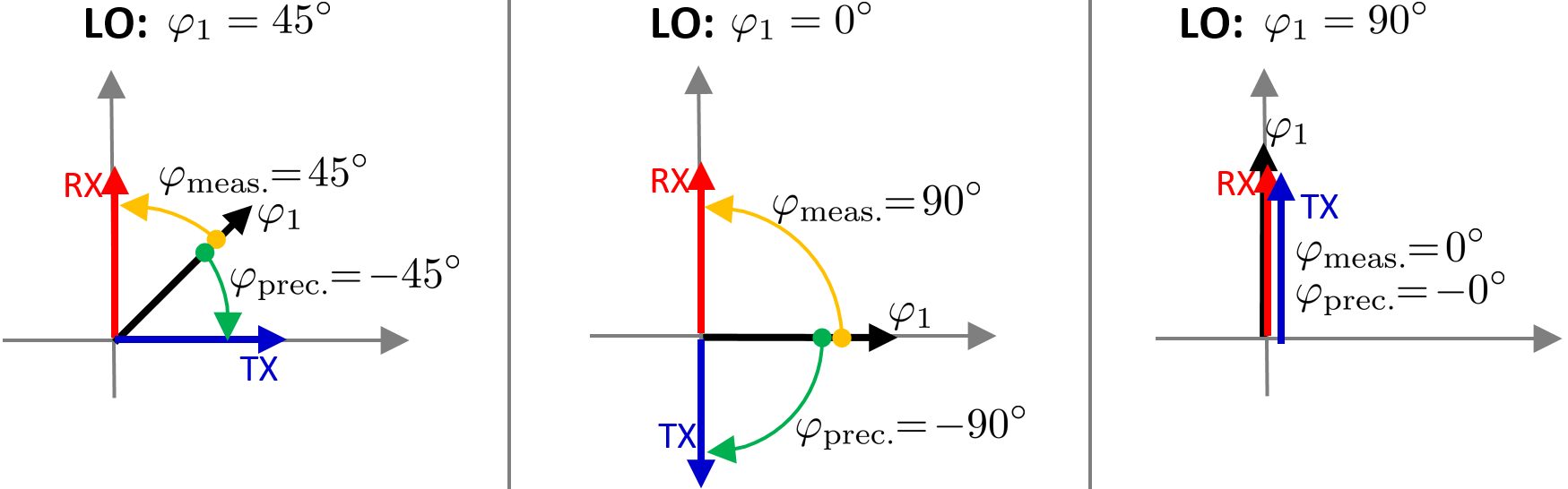}
	\caption{Depending on the LO phase, the phase of the precoded TX signal can be completely different, implying that calibration between \ac{TRP}s is of upmost importance for coherent joint transmission.}
	\label{fig:exampleLO}
\end{figure}

\noindent
Let us first consider a simple example to show how severe the \ac{LO} effect is once the correct \ac{TX} and \ac{RX} model is used. As illustrated in Figure~\ref{fig:exampleLO},
a reference \ac{RX} signal with a ``global'' phase\footnote{Relative to a global (mathematical) reference phase. This reference phase can be arbitrarily defined but not directly measured (except if the global reference coincides with one of the LOs).}
of $\varphi_\text{RX} = 90^\circ$ is assumed. One cannot measure this ``global'' phase, only $\varphi_\text{meas.} = \varphi_\text{RX} - \varphi_1 =  45^\circ$ (left picture) can be measured, i.e., relative to the \ac{LO}. 
The precoded signal is then the conjugate of the measured \ac{RX} signal,  $\varphi_\text{prec.} = -\varphi_\text{meas.} =  -45^\circ$. When sending out this signal, it will again be relative to the \ac{LO}, i.e., the ``global'' \ac{TX} phase is  $\varphi_\text{TX} = \varphi_\text{prec.} + \varphi_1  =  0^\circ$. Suppose this is exactly the ``global'' \ac{TX} phase which adds up coherently with the signal of another \ac{TRP} at the \ac{UE}, i.e., everything is calibrated. 
If now the \ac{LO} phase of \ac{TRP}1 drifts to $\varphi_1 =  0^\circ$ (middle picture), while everything else stays constant, coherency will be lost. This can  easily be seen by repeating the same procedure as before, i.e., measure the \ac{RX} signal $\varphi_\text{meas.} = 90^\circ$ and send the conjugate of it $\varphi_\text{prec.} = -90^\circ$, leading to a ``global'' \ac{TX} phase of $\varphi_\text{TX} = -90^\circ$.  If the system was coherent before, this coherency is now broken. Even more surprising, if the old channel estimation would have been used, $\varphi_\text{meas.} = 45^\circ$, the overall \ac{TX} phase error would be smaller.  
This simple example illustrates that whenever the \ac{LO} phases drift apart, a new calibration becomes necessary.

\subsection{Relative Reciprocity Calibration: Why Free-Running LOs are Not Useful}

\begin{figure}[!t]
	\centering
	\includegraphics[width=\columnwidth]{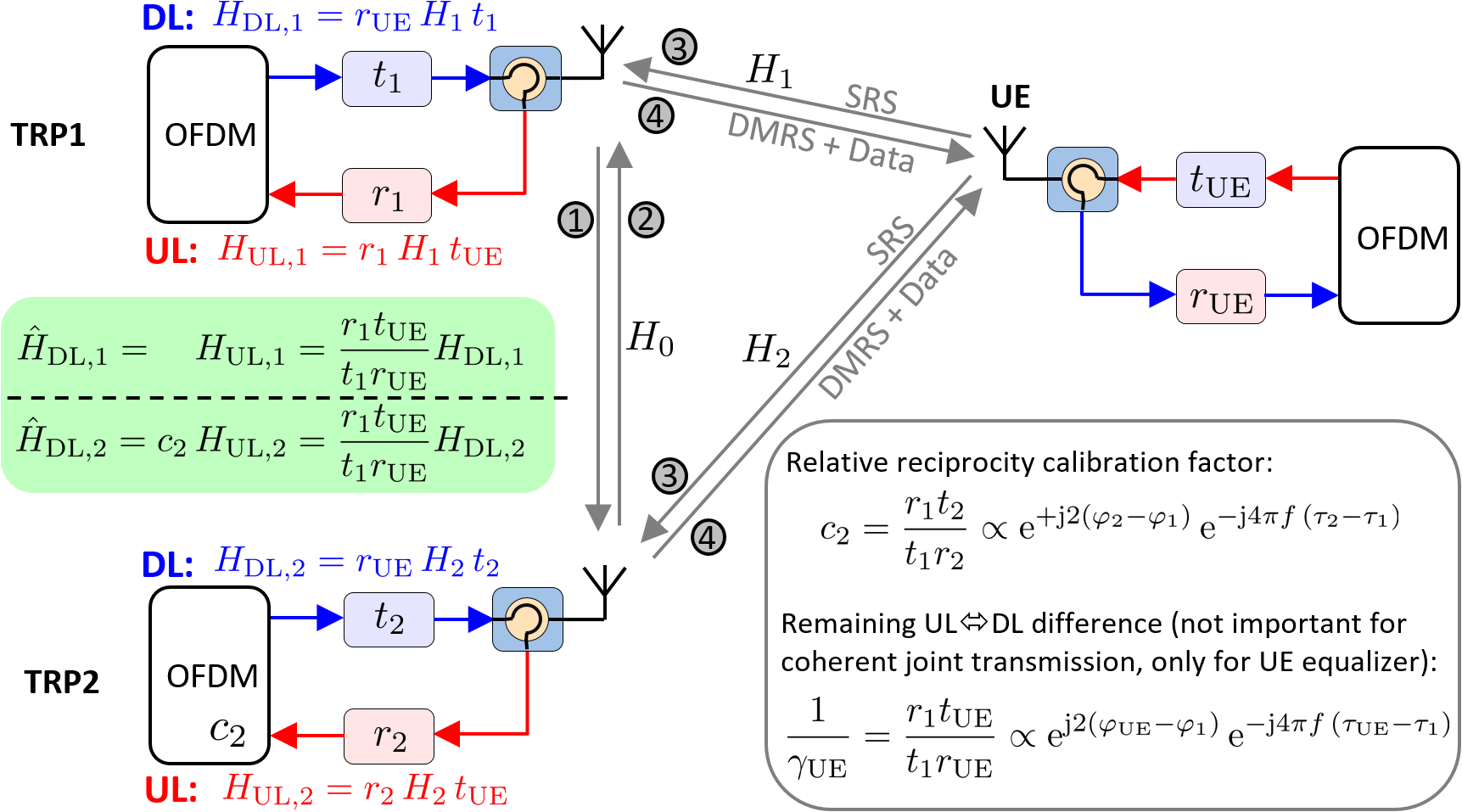}
	\caption{Concept of relative reciprocity calibration. The extension to more \ac{TRX},  \ac{TRP}s and \ac{UE}s is trivial and thus omitted.}
	\label{fig:relativeCalibration}
\end{figure}

\noindent
As already described above, coherent joint transmission requires reciprocity calibration. A practical method to achieve that is \ac{OTA} relative reciprocity calibration \cite{shepard2012argos}, where one \ac{TRX} acts as a reference. 
Without loss of generality, let us consider two \ac{TRP}s and one \ac{UE}, as  illustrated in Figure~\ref{fig:relativeCalibration}. 
By combining \eqref{eq:H_UL} and \eqref{eq:H_DL}, the \ac{UL} channel can be expressed by the \ac{DL} channel according to:
\begin{align}
	\label{eq:HDLest1}
	{\hat{H}}_\text{DL,1} = \ \ \ {H}_\text{UL,1}
	&=
	\frac{r_1 t_\text{UE}}{t_1 r_\text{UE}}
	{H}_\text{DL,1}
	\\
	{\hat{H}}_\text{DL,2} = c_2 \,{H}_\text{UL,2} &=
	\frac{r_1 t_\text{UE}}{t_1 r_\text{UE}}
	{H}_\text{DL,2}
	.
	\label{eq:HDLest2}
\end{align}
For \ac{TRP}1, the estimated \ac{DL} channel, ${\hat{H}}_\text{DL,1}$, is directly given by the \ac{UL} channel estimate (assumed to be perfectly known), i.e., this is our reference \ac{TRX}. If the same method would be applied for \ac{TRP}2, coherency cannot be guaranteed because of the different factors $\frac{r_1}{t_1}$ at \ac{TRP}1 and $\frac{r_2}{t_2}$ at \ac{TRP}2. Thus, calibration becomes necessary, i.e., we have to multiply the \ac{UL} channel estimate at \ac{TRP}2 with the relative reciprocity calibration factor $c_2 = \frac{r_{1} t_2}{ t_1 r_2}$. 
This calibration is relative in the sense that the signals of the two \ac{TRP}s will add up coherently at the \ac{UE}, but with some unknown common phase rotation $( \frac{r_1 t_\text{UE}}{t_1 r_\text{UE}} )^*$, which needs to be estimated by the \ac{UE} e.g., by \ac{DMRS}. 
Calibration factor $c_2$ can be obtained by sending a signal from \ac{TRP}1 to \ac{TRP}2 in time-slot 1, delivering measurement $y_{2,1}[1]$. In a second time-slot, \ac{TRP}2 will send a calibration signal to \ac{TRP}1, delivering measurement $y_{1,2}[2]$:
\begin{align}
	y_{2,1}[1] &= r_2 \, H_{0} \, t_1
	\\
	y_{1,2}[2] &= r_1\, H_{0} \, t_2	
.	
\end{align}
By dividing those measurement results, the relative reciprocity calibration factor $c_2$ becomes, 
\begin{align}
	\label{eq:c2_a}
		c_2(f) &= \frac{y_{1,2}[2]}{y_{2,1}[1]} = \frac{r_{1} H_0 t_2}{t_1 H_0 r_2 }  = \frac{r_{1} t_2}{t_1 r_2 }
		 \\
		 &= 
		 \mathrm{e}^{+ \mathrm{j}2( \varphi_{2}-\varphi_{1})}
		 \,
		 \mathrm{e}^{ -\mathrm{j} 4 \pi f \, (\tau_2 - \tau_1)}
		 \frac{
		 	H_{r,1}(f)\,H_{t,2}(f)
		 }
		 {
		 	H_{r,2}(f)\,H_{t,1}(f)
		 }.		
	 \label{eq:c2} 
\end{align}
Note that a stable channel and a stable \ac{HW} is assumed in \eqref{eq:c2_a}, the more general solution would include time variations of $r$, $t$ and $H_0$, but is omitted here for the sake of clarity.

The implications of \eqref{eq:c2} are significant.
\textbf{As soon as the \ac{LO} phases $\varphi_{2}-\varphi_{1}$ drift apart, a new reciprocity calibration\footnote{
		There exist alternative methods to estimate the new calibration factor based on \ac{SRS}, but they are rather complicated and quite scenario and setup dependent (i.e., will not always work).} becomes necessary, otherwise coherency is lost.}  
In the most straightforward approach, the number of \ac{OTA} calibration signals is proportional to the number of uncalibrated \ac{TRX} \cite{shepard2012argos}. 
While this simple approach can be improved, e.g., signals can be frequency multiplexed or avalanche-like methods can be employed \cite{jiang2018framework}, the calibration overhead for free-running \ac{LO}s is still very large and completely unnecessary. 
Simply by locking the \ac{LO} phases
to some common reference clock, most of the problems can be avoided. Thus, the clear recommendation is to lock \ac{LO}s as much as possible.
In a distributed \ac{MIMO} setup, locking \ac{LO}s is of course more challenging but one can still distribute a common clock signal over cable (e.g., clock recovery). However, time variations of the long cable (e.g., because of temperature variations) will also cause \ac{LO} phase drifts, implying that reciprocity calibration needs to be performed more frequently than in a pure massive \ac{MIMO} system, but still less often than for free-running \ac{LO}s. 

Note that in this paper timing errors $\tau_2 - \tau_1$  will be ignored since they typically have less impact on phase errors when compared with the \ac{LO}, i.e., \ac{LF} vs. \ac{RF}, but of course this depends on \ac{HW} implementation and could also be a major problem.

\subsection{Perfect Reciprocity Calibration: A Problematic Assumption}

\noindent
A common assumption in literature is perfect reciprocity calibration, i.e., the \ac{DL} channel can be perfectly estimated from the \ac{UL} channel (ignoring noise and interference). Considering \eqref{eq:HDLest1} and \eqref{eq:HDLest2}, this can be achieved if all \ac{TRP}s know \ac{UE} reciprocity calibration factor $\gamma_{\text{UE}}$, so that
\begin{align}
	\label{eq:HDLest1b}
	{\hat{H}}_\text{DL,1} = \ \ \   \gamma_{\text{UE}}  \, {H}_\text{UL,1}
	&=
	{H}_\text{DL,1}
	\\
	{\hat{H}}_\text{DL,2} = \gamma_{\text{UE}} \, c_2 \,{H}_\text{UL,2} &=
	{H}_\text{DL,2}
	,
	\label{eq:HDLest2b}
\end{align}
with
\begin{align}
	\label{eq:gamma}
	 \gamma_{\text{UE}}(f) &=
	\frac{t_1 r_\text{UE}}{
		r_1 t_\text{UE}
	}
\\
&=
		 \mathrm{e}^{- \mathrm{j}2( \varphi_{\text{UE}}-\varphi_{1})}
\,
\mathrm{e}^{ \mathrm{j} 4 \pi f \, (\tau_\text{UE} - \tau_1)}
\frac{
	H_{r,\text{UE}}(f)\,H_{t,1}(f)
}
{
	H_{r,1}(f)\,H_{t,\text{UE}}(f)
}	
		\label{eq:gamma2}
\end{align}
Similar to the relative reciprocity calibration factor, the implications of \eqref{eq:gamma2} are severe.  
\textbf{As soon as the \ac{LO} phases} $\varphi_{\text{UE}}-\varphi_{1}$ \textbf{drift apart, perfect reciprocity calibration is lost and simple blind \ac{DL} channel estimation no longer works}. 
Sending more \ac{SRS} symbols also does not help\footnotemark[3], it actually worsens the performance. 
Thus, one needs to measure the \ac{UE} reciprocity calibration factor again,
but this is usually quite cumbersome: Firstly, the \ac{UE} needs to measure the phase of the effective \ac{DL} channel by \ac{DMRS}, i.e., $ \arg \{1/\gamma_{\text{UE}}^*\} = \arg \{\gamma_{\text{UE}}\}$. Secondly, it needs to feedback this information to the \ac{TRP}s, wasting \ac{UL} resources. Finally, the \ac{TRP} can change the \ac{TX} phase so that the signal arrives at the \ac{UE} with phase zero (assuming everything was stable enough so that the feedback delay has no influence).
However, if the \ac{UE} has already estimated the effective \ac{DL} channel, it can use this information directly for equalization and there is simply no point in sending back this information. Only in the unrealistic case of a very stable $\varphi_{\text{UE}}-\varphi_{1}$, perfect reciprocity calibration might have some advantages in some scenarios, but considering the very good alternative of relative reciprocity calibration together with low overhead and robust \ac{DMRS}, which not only track the \ac{LO} phases but also time variations of the propagation channel, the whole concept of perfect reciprocity calibration seems not very useful.

\subsection{SRS and DMRS}

\noindent
\ac{SRS} and \ac{DMRS} have been mentioned several times, so let us quickly look at those reference signals. 
The purpose of \ac{SRS} is to estimate the full \ac{UL} channel matrix, i.e., $\mathbf{H}_\text{UL} \in \mathbb{C}^{N_\text{TRX} \times N_\text{UEs}}$, and requires a minimum of $N_\text{UEs}$ orthogonal time-frequency resources, though sending more \ac{SRS} and averaging out noise and interference can be advantageous.
\ac{DMRS}, on the other hand, are intermixed within the \ac{QAM} data symbols (they use the same precoder) and allow to estimate the effective \ac{DL} channel at the \ac{UE}, $\mathbf{a} = \text{diag} \{\mathbf{H}_\text{DL} \mathbf{W}\} \in \mathbb{C}^{N_\text{UEs} \times 1}$, with $\mathbf{W} \in \mathbb{C}^{N_\text{TRX} \times N_\text{UEs}}$ denoting the precoder. A minimum of only one time-frequency resource is required to estimate all elements of $\mathbf{a}$, though sending more \ac{DMRS} and averaging out noise and interference can again be advantageous.
For example, spending only one time-frequency position for \ac{DMRS}, there is a channel estimation penalty\footnote{$\text{SINR}_\text{eff.} = \frac{|\mathcal{P}|\text{SINR}}{|\mathcal{P}|+1+1/\text{SINR}} $, as analytically derived in \cite[Equation (4.27)]{NisselDissertation} for single antenna BEP in Rayleigh fading (after some reformulations). Throughput simulations show a similar behavior in massive MIMO.} of roughly 3dB (1+1/1) on the \ac{SINR}, while for averaging over six time-frequency position, as used in 5G as one possible configuration (within 12 subcarriers and 14 OFDM symbols), the penalty is only 0.7dB (1+1/6). Thus, the overhead of \ac{DMRS} is very low, while the performance very good, explaining why they are so useful.

\section{Numerical Example}

\begin{figure}[!t]
	\centering
	\includegraphics[width=\columnwidth]{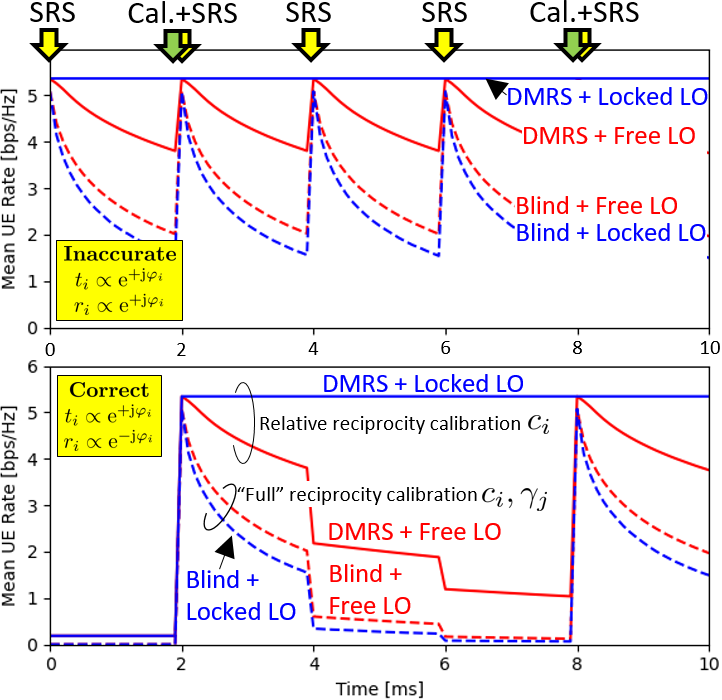}
	\caption{The correct TX and RX model leads to completely different conclusions when compared with the inaccurate model. All LOs used for coherent joint transmission should be locked. Moreover, relative reciprocity calibration together with DMRS offer a very good performance. 
}
	\label{fig:numericalExample}
\end{figure}

\noindent
The numerical example shown in Figure~\ref{fig:numericalExample} is based on a similar setup as \cite{bjornson2015distributed}. In total there are 16\,\ac{TRP}s, each with 64\,\ac{TRX}s. Four \ac{TRP}s form a distributed massive \ac{MIMO} cluster and the total network consists of $10\times16=160$\,\ac{UE}s. The phase noise model is the same as in \cite{mehrpouyan2012joint}, i.e., $\varphi[n] = \varphi[n-1] + \Delta$, with $\Delta \sim \mathcal{N}(0,\sigma^2=0.01)$ and $n$ representing the time index (sampling time $T_\text{S} = 100\,\text{µs}$). Moreover, zero-forcing precoding is assumed, the \ac{LO} phases have a random initial value, no other \ac{HW} effects are present, i.e., $\tau_i=0, H_{t,i}=H_{r,i} = 1$, and perfect channel estimation (\ac{SRS} and \ac{DMRS}) is available. When using the inaccurate \ac{TRX} model, reciprocity calibration is not needed because the \ac{LO} phases become part of the channel and will be implicitly estimated by \ac{SRS}. On the other hand, with the correct model, reciprocity calibration is of upmost importance. 
Free-running \ac{LO}s and blind \ac{DL} channel estimation are both not very useful because they would require continuous calibration signals and have overall a very poor performance.  
Instead, all \ac{LO}s used for coherent joint transmission should be locked and   relative reciprocity calibration together with \ac{DMRS} employed.

\section{Conclusions}

\noindent

\noindent
Some common misunderstandings related to reciprocity calibration have been clarified in this paper. 
Essentially, coherent joint transmission can only work if the \ac{LO} phases are locked, either directly by ``cables'' or indirectly by continuously sending reciprocity calibration signals. The first method is typically preferred as it does not ``waste'' any time-frequency resources.
The commonly made assumption of perfect reciprocity calibration typically does not hold, as it would require too much calibration overhead. 
Instead, relative reciprocity calibration is a better option.
Thus, to make blind  \ac{DL} channel estimation at the \ac{UE} practically useful, one must also be able to blindly estimate the phase, mostly ignored in literature so far. Alternatively, by sending low-overhead \ac{DMRS} many problems can be avoided and is therefore the recommended solution for effective \ac{DL} channel estimation at the \ac{UE}.

{
	\appendix

\begin{figure}[!t]
	\centering
	\includegraphics[width=\columnwidth]{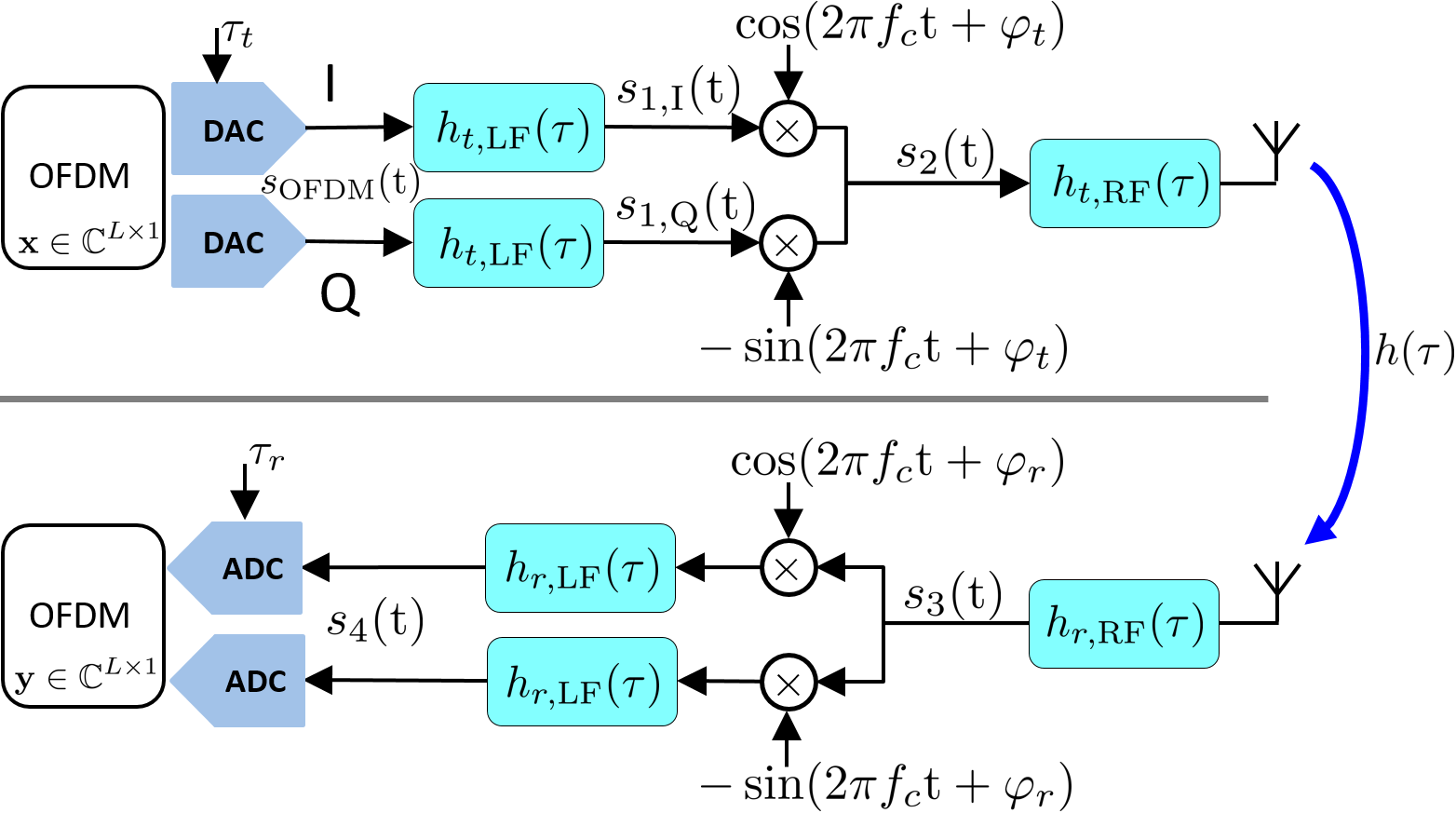}
	\caption{Block diagram of a generic transmission chain, used to derive the TX and RX transfer functions $t_i$ and $r_i$. }
	\label{fig:BlockDiagramtXRX}
\end{figure}

\noindent
Figure~\ref{fig:BlockDiagramtXRX} shows a generic transmission chain, used to mathematically proof the simplified \ac{TX} and \ac{RX} model in \eqref{eq:tx} and \eqref{eq:rx}. Note that all elements in Figure~\ref{fig:BlockDiagramtXRX} are assumed to be perfect in order to focus on the key aspects, i.e., perfectly linear, no IQ imbalances, no phase noise within one \ac{OFDM} symbol, etc.
The transmitted \ac{OFDM} signal in the time domain, $s_\text{OFDM}(\text{t})$, can be written as \cite{NisselDissertation},
\begin{align}
	\label{eq:OFDM}
	 s_\text{OFDM}(\text{t}) = \sum_{l= -\frac{L}{2} }^{ \frac{L}{2}-1} \mathrm{e}^{\mathrm{j}2\pi \, l F \,  (\text{t} -\tau_t)} \,x_l ,
\end{align}
where $x_l$ denotes the transmitted \ac{QAM} data symbol at subcarrier position $l$, $F$ the subcarrier spacing, $L$ the total number of subcarriers, and $\tau_t$ the delay of starting the transmission (vs. a perfect ``global'' time reference).
Note that the rectangular window function of \ac{OFDM} is ignored in \eqref{eq:OFDM}, which is fine as long as the cyclic prefix is sufficiently long. 
The I-signal experiences the real valued impulse response $h_{t,\text{LF}}(\tau)$, leading to,
\begin{align}
	\label{eq:fristStage}
\begin{split}
s_{1,\text{I}}(\text{t})
 &= \Re\{ s_\text{OFDM}(\text{t}) \}  \ast h_{t,\text{LF}}(\text{t})
 \\   
 &=	
\Re
 \Big \{
  \sum_{l=-\frac{L}{2}}^{ \frac{L}{2}-1} \mathrm{e}^{\mathrm{j}2\pi \, l F \,  (\text{t} -\tau_t)} x_l 
  \underbrace{
  \int\limits_{-\infty}^{\infty}  h_{t,\text{LF}}(\tau) \mathrm{e}^{-\mathrm{j}2\pi \, l F \, \tau} \mathrm{d}\tau
  }_{ H_{t,\text{LF}}(lF)}
  \Big \}
\end{split}
,
\end{align}
with $H_{t,\text{LF}}(f)$ denoting the transfer function of $h_{t,\text{LF}}(\tau)$, i.e., its Fourier transform. 
The Q-signal, $s_{1,\text{Q}}(\text{t})$, is similar except for taking the imaginary part instead of the real part. 
Utilizing Euler's formula, the up-converted signal can be written as:
\begin{align}
	\begin{split}
	s_2(\text{t}) &= s_{1,\text{I}}(\text{t}) \cos(2 \pi f_c \text{t}+ \varphi_t) - s_{1,\text{Q}}(\text{t}) \sin(2 \pi f_c \text{t}+ \varphi_t)
	\\
	& = 
	\Re \Big  \{ \mathrm{e}^{\mathrm{j} \varphi_t }
	  \sum_{l=-\frac{L}{2}}^{ \frac{L}{2}-1} \mathrm{e}^{\mathrm{j}2\pi (l F + f_c) \text{t}} \, \mathrm{e}^{-\mathrm{j}2\pi l F \tau_t}  \,H_{t,\text{LF}}(lF) \, x_l   	
	  \Big \},
	\end{split}
\end{align}
with $f_c$ denoting the carrier frequency. 
The convolution of $s_2(\text{t})$, \ac{TX} \ac{RF} chain $h_{t,\text{RF}}(\tau)$, propagation channel $h(\tau)$ and \ac{RX} \ac{RF} chain $h_{r,\text{RF}}(\tau)$ leads to $s_3(\text{t}) = s_2(\text{t})  \ast h_{t,\text{RF}}(\text{t})  \ast h(\text{t})  \ast h_{r,\text{RF}}(\text{t})$, which can be similarly described as in \eqref{eq:fristStage}, i.e., using their Fourier transforms.
Again, employing Euler's formula, the baseband signal $s_4(\text{t})$ can be written as,
\begin{align}
	\begin{split}
	s_4(\text{t}) &= h_{r,\text{LF}}(\text{t})  \ast  s_3(t) [\cos(2 \pi f_c \text{t} + \varphi_r) - \mathrm{j} \sin(2 \pi f_c \text{t} + \varphi_r)]
	\\
	&= \frac{1}{2}
	\mathrm{e}^{\mathrm{j} \varphi_t } \mathrm{e}^{-\mathrm{j} \varphi_r }
	\sum_{l=-\frac{L}{2}}^{ \frac{L}{2}-1} \mathrm{e}^{\mathrm{j}2\pi l F \, \text{t}} \, \mathrm{e}^{-\mathrm{j}2\pi l F \tau_t}  \,H_{t,\text{LF}}(lF) \\ & \quad \    H_{t,\text{RF}}(lF+f_c) \, H(lF+f_c) \,  H_{r,\text{RF}}(lF+f_c) \\ & \quad \ H_{r,\text{LF}}(lF) \, x_l  \, +  \frac{1}{2} \mathrm{e}^{- \mathrm{j} 4\pi f_c \text{t}} z(\text{t})
	,
	\end{split}
\end{align}
where $z(\text{t})$ is not relevant because it will be averaged out later. Finally, received data symbol $y_l$ can be obtained by \ac{OFDM} demodulation (assumed to be delayed by $\tau_r$) according to,
\begin{align}
	\begin{split}
	y_{l} &= 
	2F \! \! \!
	\int\limits_{\tau_r}^{\tau_r + \frac{1}{F}} 
	\! \! \!
s_4(\text{t})
 \mathrm{e}^{- \mathrm{j}2\pi \, l F \, (\text{t} - \tau_r)}
	\mathrm{d}\text{t}
	\\
	& = r(lF) \, H(lF+f_c) \,	t(lF) \, x_l
	,
\end{split}
\end{align}
with
\begin{align}
	\label{eq:tfL}
	t(lF) &= \mathrm{e}^{\mathrm{j} \varphi_t } \,   \mathrm{e}^{-\mathrm{j}2\pi l F \tau_t} \, H_{t,\text{LF}}(lF) \,   H_{t,\text{RF}}(lF+f_c)
	\\
	r(lF) &= \mathrm{e}^{-\mathrm{j} \varphi_r }  \,\mathrm{e}^{\mathrm{j}2\pi l F \tau_r}
	 H_{r,\text{LF}}(lF) \,   H_{r,\text{RF}}(lF+f_c)
	 	\label{eq:rfL}
	 	.
\end{align}
Within one \ac{TRX}, the \ac{LO} phases and time delays will be derived from the same common reference,
\begin{align}
	\label{eq:tauphi}
\begin{matrix}
	\varphi_t =  \varphi_i + \Delta\varphi_t &  & & & \tau_t =  \tau_i + \Delta\tau_t
	\\
	\varphi_r =  \varphi_i + \Delta\varphi_r & & & & \tau_r =  \tau_i + \Delta\tau_r	
\end{matrix}
.
\end{align}
Combining \eqref{eq:tfL}, \eqref{eq:rfL} and \eqref{eq:tauphi} directly leads to $t_i(f)$ and $r_i(f)$, presented in the main text, see \eqref{eq:tx} and \eqref{eq:rx}.

}

\bibliographystyle{IEEEtran}
\bibliography{bibliography}

\end{document}